\documentclass[a4paper]{icrc}  

\usepackage{times}

\usepackage[dvips]{graphicx} 

\usepackage{amsmath}
\usepackage{amssymb}
\usepackage{mathptmx}

\newcommand{\um}[1]{\ensuremath{\,\mathrm{#1}}} 

\begin{document}

\title{The AMS Time of Flight System}

\author[1]{L. Baldini}
\affil[1]{INFN, Bologna, Italy}
\author[2]{L. Brocco}
\author[2]{D. Casadei}
\affil[2]{Bologna University and INFN, Bologna, Italy}
\author[3]{G. Castellini}
\affil[3]{CNR-IROE, Florence, Italy}
\author[1]{F. Cindolo}
\author[2]{A. Contin}
\author[1]{G. Laurenti}
\author[2]{G. Levi}
\author[2]{F. Palmonari}
\author[2]{A. Zichichi}

\correspondence{D. Casadei, Diego.Casadei@bo.infn.it}

\runninghead{Baldini \emph{et al.}: The AMS TOF System}
\firstpage{2211}
\pubyear{2001}

\maketitle

\begin{abstract}
The Time of Flight (TOF) system of the AMS experiment provides the
fast trigger to the detector and measures the crossing particle
direction, velocity and charge.  AMS was operated aboard of the
shuttle Discovery on June 1998 (NASA STS-91 mission) and will be
upgraded and installed on the International Space Station at the end
of 2003, for 3 years of data taking.  The performances of the TOF
during the precursor flight and modifications needed in the final version
of the detector are presented.
\end{abstract}

\section{Introduction}

The \emph{Alpha Magnetic Spectrometer} (AMS) \citep{amsfirst} is a
particle detector that will be installed on the International Space
Station in 2003 to measure cosmic ray fluxes for at least three
years.

\begin{figure}[t]\centering
\includegraphics[width=\columnwidth]{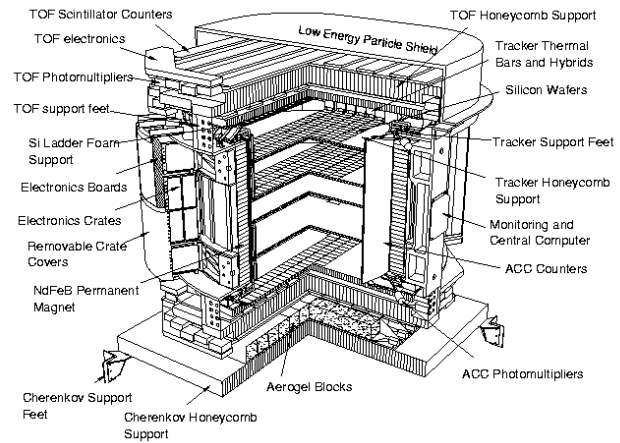}
\caption{The AMS detector for the STS-91 mission (AMS-1).}\label{AMS}
\end{figure}

During the precursor flight aboard of the shuttle Discovery (NASA
STS-91 mission, 2--12 June 1998), AMS collected data for about 180
hours \citep{amsall}.  Figure~\ref{AMS} shows the detector (called
AMS-1 in the following), consisting of a permanent Nd-Fe-B magnet, six
silicon tracker planes, an anticoincidence scintillator counter
system, the time of flight (TOF) system consisting in four layers of
scintillator counters and a threshold aerogel \v{C}erenkov detector.

The TOF system \citep{tof1} was completely designed and built at the
INFN Laboratories in Bologna. Its main goals are to provide the fast
trigger to AMS readout electronics, and to measure the particle
velocity ($\beta$), direction, position and charge.  In addition, it
had to operate in space with severe limits for weight and power
consuption.

Each TOF plane consists of 14 scintillator counters covering a roughly
circular area of 1.6\um{m^2}.  The scintillation light is guided to 3
Hamamatsu R5900 photomultipliers per side, whose signals are summed to
have a good redundancy and light collection efficiency.  The total
power consumption of the system (112 channels, 336 phototubes) was
150\um{W}, while its weight (support structure included) was
250\um{kg}.

\begin{figure}[t]\centering
\includegraphics[width=\columnwidth]{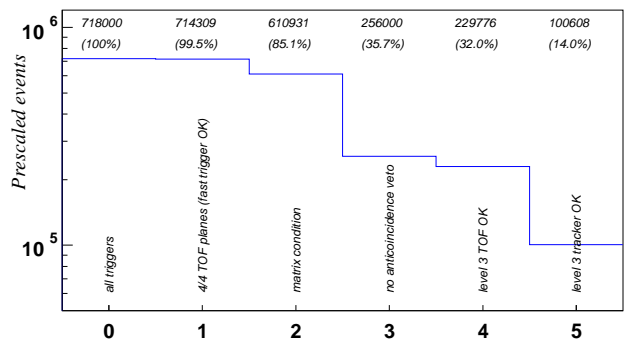}
\caption{Trigger cuts on prescaled events.}\label{trigger}
\end{figure}

\section{The AMS-1 trigger}
The AMS-1 trigger logic consists of three levels.  The \emph{fast
trigger} (FT) processes the analog scintillators data and provides, in
about 50\um{ns}, the zero time for the time-of-flight measurement. The
\emph{first level trigger} rejects events with hits on the
anticoincidence counter system and enhances the fraction of particles
crossing the tracker planes through the analysis of the pattern of hit
counters in the first and fourth TOF plane.  The \emph{second level
trigger} (nicknamed ``third level trigger'' for historycal reasons)
refines the TOF trigger and finds preliminary tracks on the silicon
tracker by using the digitized data.

Figure \ref{trigger} shows the reduction given by the different
trigger conditions on a subsample of data taken requiring only the
fast trigger (\emph{prescaled events}) in the ratio of 1 to 1000
normal triggers.

\begin{figure}[t]\centering
\includegraphics[width=\columnwidth]{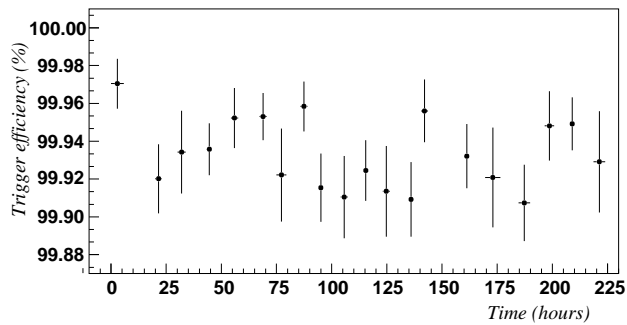}
\caption{Fast trigger efficiency during the STS-91 shuttle
flight.}\label{trig-eff-time}
\end{figure}

\subsection{Off-time events}

The FT signal is generated when at least one counter side in each of
the TOF planes produces a signal above a threshold corresponding to
60\% of a minimum ionizing particle.  To measure the efficiency of
this first event selection an unbiased sample of events is required.

This was done by exploiting the characteristics of the TOF read-out
electronics to be sensitive to particles impinging in the detector in
an interval of about 16\um{\mu s} around the trigger signal.  Both
the (un-discriminated) amplitude and the discriminated output of up to
eight hits can be registered by each channel with a time resolution of
1\um{ns}.

In this way \emph{off-time} particles can be reconstructed in a
totally unbiased way.  On average, there are $0.05\div0.7$ off-time
events per trigger, depending on the trigger rate.

\begin{figure}[t]\centering
\includegraphics[width=\columnwidth]{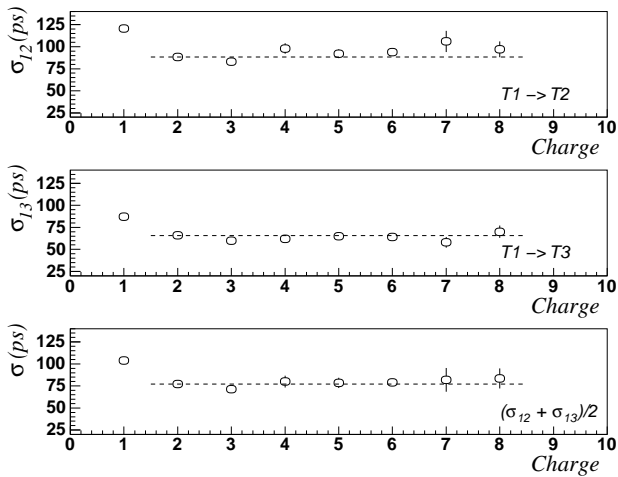}
\caption{Single plane time resolution from the time of flight between
the first and the second ($\sigma_{12}$) or the third ($\sigma_{13}$)
TOF plane, and the mean time resolution.}\label{timeres}
\end{figure}
\begin{figure}[t]\centering
\includegraphics[width=\columnwidth]{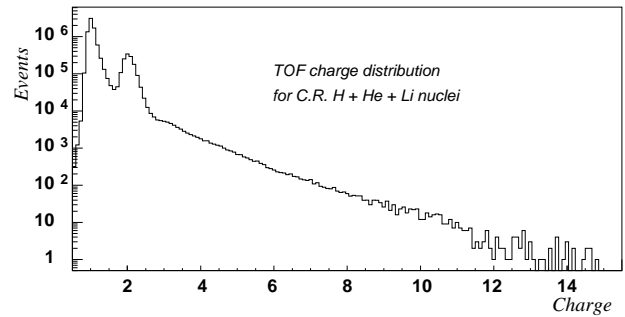}
\caption{TOF charge distribution for $Z<4$
nuclei seen by the AMS-1 detector during the STS-91 flight.}\label{tof-part-sep}
\end{figure}
\begin{figure}[h!]\centering
\includegraphics[width=\columnwidth]{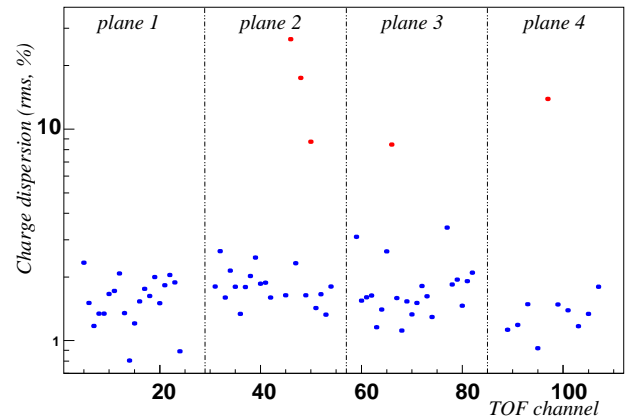}
\caption{TOF charge peaks dispersion during the STS-91
flight.}\label{charge-rms}
\end{figure}

\subsection{Fast trigger efficiency and background}

The fast trigger efficiency can be derived as the ratio between the
number of all off-time particles and the number of off-time particles
which produce signals greater than the threshold in all counters
hit. Figure~\ref{trig-eff-time} shows that this efficiency, for the
whole duration of the STS-91 flight\footnote{Data taken in the South
Atlantic Anomaly are excluded from the analysis.}, was always above
99.9\%.

The background can be estimated by analysing the prescaled events.
The ratio between bin 1 and bin 0 of figure~\ref{trigger} shows that
only about 0.5\% of the fast triggers were due to noise, but this
background is completely eliminated by requiring the coincidence of
both sides of the same counter in the third level trigger.

\section{Time of flight resolution}

The single channel time resolution is \citep{tof1}:
\[
   \sigma(x) = \sqrt{ \frac{\sigma_1^2}{N} + \frac{\sigma_2^2 x^2}{N}
		+ \sigma_3^2 } \; ,
\]
where $x$ is the distance of the particle crossing point from the
photomultiplier (PM), $N$ is the number of photons which convert on
the PM window, $\sigma_1$ depends upon the PM signal shape and the
trigger electronics, $\sigma_2$ takes into account the dispersion in
the photons path lengths and $\sigma_3$ is the electronic noise.

The overall time resolution of a plane can be determined by measuring
the time of flight of ultrarelativistic particles between two given
planes, after correcting for the track length.

The time dispersion is expected to decrease with the nuclear charge
$Z$, due to the large number of photoelectrons produced by nuclei with
high atomic number, until it reaches the minimum value $\sigma_3$,
dictated by the electronic noise.  Figure~\ref{timeres} shows the time
resolution measured using the first and the second or the first and
the third TOF plane: the horizontal lines show that the level of the
electronic noise is 88\um{ps} for the first measurement and 66\um{ps}
for the second one, leading to a mean value of $\sigma_3 =77 \um{ps}$,
which represents the limiting resolution of the system.

\section{Photomultipliers stability}

The TOF system provides a measurement of the absolute charge of the
crossing particle in addition to the tracker.

Due to the strong constraints about power consuption, the TOF
front-end electronics was not optimized for energy deposition
measurements. The signal from the PM anodes was integrated and
discriminated with a threshold (on the integrated signal) set to about
20\% of the minimum ionizing particle.  The resulting
time-over-threshold is proportional to the logarithm of the deposited
charge.  The method results in a good separating power between charges
$|Z|=e$ and $|Z|>e$ (where $e$ is the proton charge) but a poor charge
resolution for $|Z|>2e$ (see figure~\ref{tof-part-sep}).

The stability of the charge measurement was very good for all the 112
TOF channels, but five channels, as shown in figure~\ref{charge-rms}.

\section{Particle separation}

At the trigger level, one goal of the TOF system was to provide a
special flag for ions.  Accordingly, it was designed to distinguish in
a fast and efficient way protons from ``$Z>e$'' particles.

Even though the low power read-out electronics was not optimized for a
good charge measurement, the TOF system energy resolution is
sufficient to separate singly by doubly charged particles with a
contamination at the level of 1\%, as shown in
figure~\ref{tof-part-sep}.

One of the main purpose of the TOF system is the measurement of the
time of flight of the particles traversing the detector with a
resolution sufficient to distinguish upward from downward going
particles: an ``upward-going'' Helium nucleus wrongly labelled
``downward-going'' would be interpreted as an
``downward-going''anti-Helium nucleus.

The average time of flight of the particles which traverse AMS is of
the order of 5\um{ns}, while the time measurement has a resolution
$\sigma_t \lesssim 120\um{ps}$, independent from the rigidity.  Thus
the probability to mistake the particle direction is well below
$10^{-11}$, the level needed for successful operation aboard the ISS,
where AMS is expected to collect at least $10^{10}$ events.

In addition, the velocity resolution of the TOF system, $\sigma
(\beta)/\beta\approx 3\%$, allows to discriminate between protons and
electrons up to a rigidity of 1.5\um{GV}.

\begin{figure}[t]\centering
\includegraphics[width=\columnwidth]{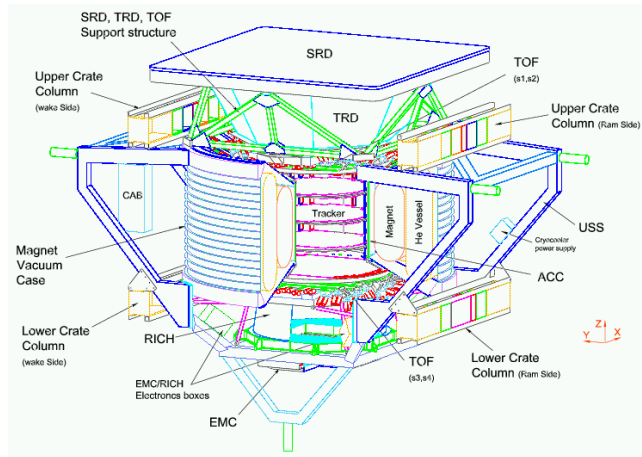}
\caption{The final version (AMS-2) of the detector.}\label{ams2}
\end{figure}
\begin{figure}[t!]\centering
\includegraphics[width=\columnwidth]{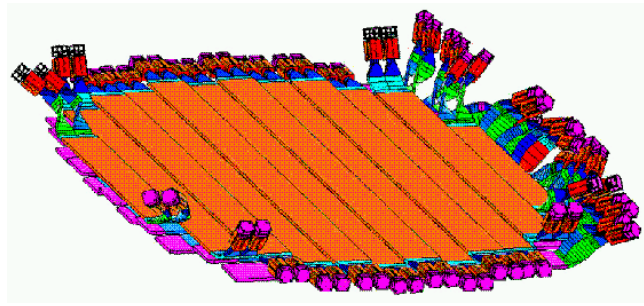}
\caption{Upper 2 planes of the new TOF system.  The PMs of plane 2 are
shown only on one side for clarity.}\label{tof2}
\end{figure}
\begin{figure}[t]\centering
\includegraphics[width=\columnwidth]{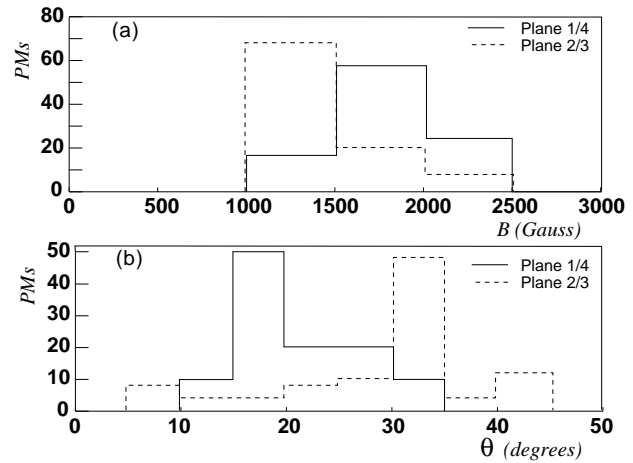}
\caption{Magnetic field intensity (a) and angle between B-field and
PM axis (b).}\label{tof2ang}
\end{figure}

\section{The time-of-flight system of AMS-2}
 
After the successful operation of AMS-1, the detector has been
redesigned to increase the maximum detectable rigidity up to $1
\um{TV}$, by using a superconducting magnet which will provide a
maximum field of about $0.8 \um{T}$.  Figure~\ref{ams2} shows the new
version of the AMS detector (called AMS-2) that will be installed on
the ISS at the end of year 2003.

In AMS-1, Hamamatsu R5900 photomultiplier tubes were used as light
detectors in that they provided small occupancy, low power consumption
and good time resolution.  In order to shield the tubes from the
residual magnetic field ($200 \um{G}$) the PMs were enclosed in a
$0.5 \um{mm}$ thick shielding case made of VACOFLUX permalloy.

The AMS-2 superconducting magnet produces a much larger field (about
$2 \div 3 \um{kG}$) of variable direction on the TOF planes.  To work
in such conditions the PMs must withstand the magnetic field without
shielding, in a large interval of angles between their axis and the
field direction.  The design of the TOF system for AMS-2 was
therefore completely determined by the choice of the light detectors.

After a market study, the Hamamatsu R5946 photomultiplier tube was
considered as the best choice and throughfully tested for time
resolution and pulse height response in magnetic field \citep{tof5}.
The results show that for angles between the PM axis and the magnetic
field greater than about $40^\circ$ the time resolution of the PMs
becomes unacceptably high.

The design of the new TOF counters is different from AMS-1 in the
following points: a) due to their larger size, only two PMs are
accomodated in each side of the counter; b) the light guides are
designed so as they can be tilted to various directions; c) due to
mechanical constraints, some of the counters have a clear plastic
extension between the scintillator and the light guides.

Each of the four TOF planes consists of 12 counters, $12\um{cm}$ wide
instead of 14 counters ($11\um{cm}$ wide) as in AMS-1.  Also, the PM
orientations cannot be completely optimized. Figure~\ref{tof2} shows
the design of planes 1 and 2 of the TOF system. The PM orientation
with respect to the magnetic field and the field magnitude are shown
in figure~\ref{tof2ang} for all PMs of the system.

The read-out electronics presently under design will be similar to
AMS-1 as for the time measurement, while the deposited charge will be
digitized with linear ADCs in order to reach a better charge
resolution.

The AMS-2 TOF system will have a worst time resolution then in AMS-1
($150\div170$\um{ps} instead of 120\um{ps}), due to the tilted light
guides and to the effect of the magnetic field. In particular, several
of the PMs will have an angle with respect to the magnetic field
direction greater than $30^{\circ}$.

\section{Conclusion}

The time of flight system for the AMS detector has proven to be a very
efficient triggering system, also capable to discriminate between
protons and heavier nuclei with a 1\% background, to measure the
particle velocity with $\sigma(\beta)/\beta\lesssim 3\%$ and the
crossing position with $\sigma(s) \lesssim 2\um{cm}$, with a total
power consumption of about $150\um{W}$ and a weight of 250\um{kg}.

The new version of the subdetector will operate with a very strong
residual magnetic field, trying nevertheless to satisfy the same
requirements and still keeping power consumption and weight at the
same level.

\end{document}